\documentstyle[aps,twocolumn,epsfig]{revtex} 
 
\begin{document} 
 
\author{Ph. H\"{a}gler$^{1}$, R. Kirschner$^{2}$, A. Sch\"{a}fer$^{1}$, 
L. Szymanowski$^{1,3}$, O.V. Teryaev$^{1,4}$} 
\address{\medskip $^{1}$Institut f\"{u}r Theoretische Physik, Universit\"{a}t 
Regensburg,
D-93040 Regensburg, Germany\\ 
$^{2}$Institut f\"{u}r Theoretische Physik, Universit\"{a}t Leipzig,
D-04109 Leipzig, Germany\\ 
$^{3}$Soltan Institute for Nuclear Studies,
Hoza 69, 00689 Warsaw, Poland\\ 
$^{4}$Bogoliubov Laboratory of Theoretical Physics, JINR, 141980 Dubna, Russia} 
\title{Direct J/$\Psi$ hadroproduction in $k_\perp$-factorization and the color
octet mechanism} 
\maketitle 
 
\begin{abstract}
The hadroproduction of direct  $J/\Psi$ in the framework of
the $k_\perp$-factorization approach is studied. The color-singlet contribution 
is essentially larger than in the collinear approach but is still 
an order of magnitude below the data. The deficit may be well described by 
the color octet contribution with the value of the matrix element 
$\langle 0|O^{J/\Psi}_8(^3S_1)|0\rangle$ substantially decreased in comparison with 
fits in the collinear factorization. This should lead to a reduction  
of the large transverse polarization, predicted in the collinear approach. 
\end{abstract} 
 
\date{04.10.2000}  
 
 
Recently we have considered the $\chi_c$ hadroproduction within the $k_\perp$ 
factorization scheme \cite{chi}. The crucial element in the description of this process  
was the effective vertex for the production of $q\bar q $ pairs with finite  
invariant mass, which appeared in calculations of the next-to-leading  
order (NLO) corrections to the Balitsky-Fadin-Kuraev-Lipatov (BFKL)  
kernel \cite{FL96}. 
 
The effective $q\bar q$-production vertex contains - apart from
the standard term -  
also additional terms describing the  
$q\bar q$ production by means of the Reggeon-Reggeon-gluon vertex  
\cite{BFKL}. 
This additional part does not contribute, if the $q\bar q$ 
pair is produced in the color-singlet (CS) state.  
For color-octet $q\bar q$ states these additional terms contribute and 
lead to a $p_\perp$ dependence which is very different from the experimentally 
observed one. Thus we concluded that 
for $\chi_c$ hadroproduction
our result suggest that the color octet mechanism (COM) is negligible. 
Specifically, a dominant color-singlet term and $k_\perp$-factorization 
provide a fair description of the data.  
 
It is interesting to extend this approach to the  
direct production of $J/\Psi$, which is known to be a long-standing 
puzzle (for review see e.g. \cite{BFY}). 

\begin{figure}[h] 
\centerline{\epsfig{file=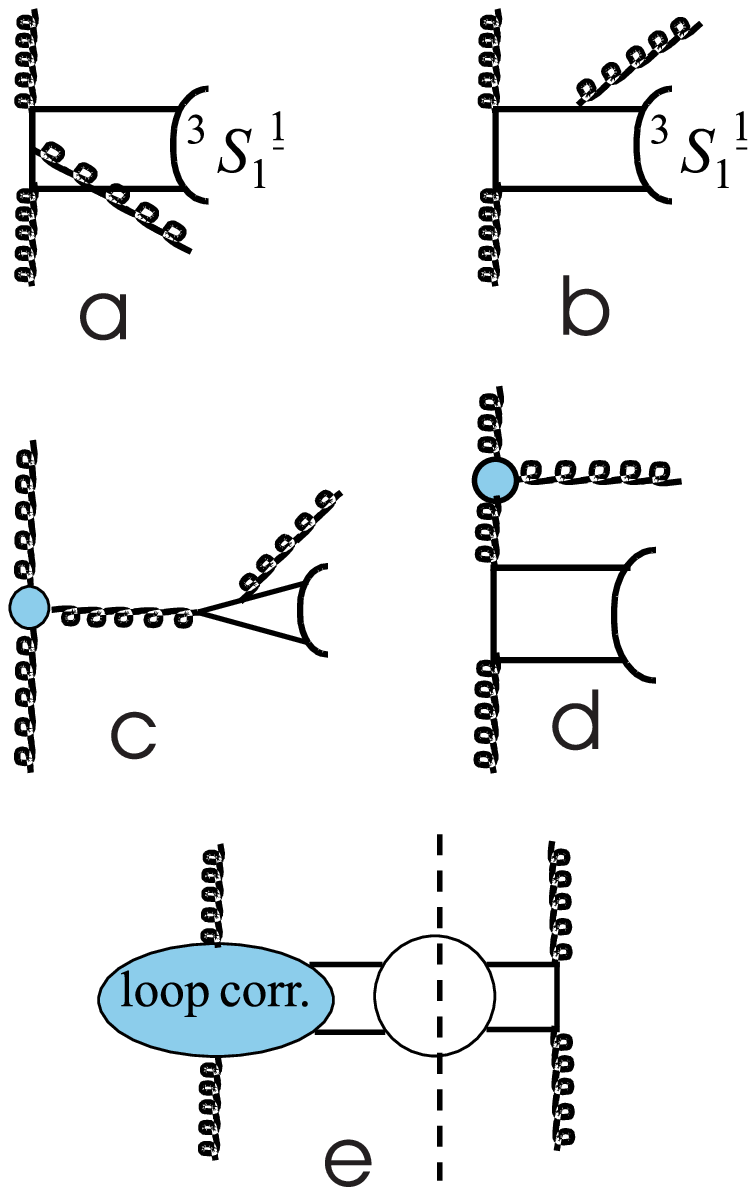,width=5cm}} 
\caption{} 
\label{graphs} 
\end{figure}

Because of the negative charge parity of the $J/\Psi$ the effective 
$q \bar q$ vertex can only give a contribution via the COM. 
In the case 
of $J/\Psi$ production in the CS model, one needs   
to know the $q\bar q$ production vertex with an additional produced
gluon (fig.\ref{graphs}). 
Such a $q\bar q\,g$ system with a finite invariant mass is called a
cluster \cite{FL96}. 
In the context of 
the BFKL approach this vertex corresponds to NNLLA corrections and is
not yet known.        
However, the unknown terms in this vertex (e.g. see figs. \ref{graphs} 
c, \ref{graphs}
d and \ref{graphs} e) should not contribute 
to the production of a CS state, in complete analogy with the $\chi_c$ 
case mentioned above \cite{chi}. 
The unknown part of the NNLLA vertex can include only diagrams 
where the additional gluon is emitted by the $t$-channel or the $s$-channel 
gluons. 
To produce a $J/\Psi$ in the CS model to LO in $\alpha_S$ 
three gluons have to be involved and one of them has to be
emitted. Only those graphs contribute, in which the emitted
gluon couples to a quark line (see fig. \ref{graphs} a and b).

 
The emission vertex of this gluon  
is described by the usual vertex since 
the interaction of partons forming  
cluster with finite 
invariant mass is governed by the usual QCD lagrangian \cite{efflagr}. 
Note that the colliding 
$t$-channel gluons are off-shell and longitudinally polarized.

In contrast to this the color-octet contribution requires the full
effective NLLA 
vertex, as 
mentioned above. It is instructive to compare this situation with the standard 
collinear factorization case \cite{CLI,CLII}. In that case, the emission of 
an additional hard gluon is required to balance the large transverse momentum 
of $J/\Psi$. The role of the color-octet states is to allow the  
fragmentation of high-$p_T$ gluons to $J/\Psi$.  
 
In our case, the transverse momentum of the colliding gluons allows for the  
production of a high-$p_\perp$ color-octet state without emission of  
an additional hard gluon. The role of the gluon fragmentation (the production  
of $J/\Psi$ by a single gluon) is now  played by the discussed additional  
term within the effective gluon vertex.

One should note the different role of singlet and octet contribution in comparison 
with $\chi_c$ production. While for $\chi_c$ the singlet and octet contributions 
are generated by the same hard scattering subprocess, the situation is different 
for $J/\Psi$. In this case the CS subprocess requires an extra gluon emission 
with respect to the CO one. Since the COM from the point of view of 
factorization corresponds to the emission of a soft gluon coupled through
non-perturbative matrix elements,  
CO and CS contributions are in the perturbative expansion formally of the same order in
$\alpha_s$. 
Therefore, one may expect a larger CO contribution
for direct
$J/\Psi$ production than for $\chi_c$ production. 
This point of view is confirmed by our results.

The  cross section for heavy quarkonium hadroproduction with
an additional produced gluon 
in the $k_{\perp}$
-factorization approach is \cite{LRSS91,RS94}, \cite{chi,open} 
\begin{eqnarray} 
&&\sigma _{P_{1}P_{2}\rightarrow J/\Psi\,g\,X} =\frac{1}{16(2\pi)^4}\!\!\int\!\!\frac{d^{3}P}{P^{+}}
\frac{d^{3}k}{k^{+}}d^{2}q_{1\perp }d^{2}q_{2\perp }
\nonumber
\\
&&\delta ^{2}(q_{1\perp }-q_{2\perp }-P_{\perp }-k_{\perp })
{\cal F}(x_{1},q_{1\perp }) 
\nonumber\\
&&\frac{1}{(q_{1\perp }^{2})^{2}} 
\left\{ \frac{\psi _{J/\Psi\,g }^{\dagger c_{2}c_{1}}\psi _{J/\Psi\,g }^{c_{2}c_{1}}
}{(N_{C}^{2}-1)^{2}}\right\} \frac{1}{(q_{2\perp }^{2})^{2}}{\cal F} 
(x_{2},q_{2\perp }).  \label{crosssection} 
\end{eqnarray}
The momenta of the $J/\Psi$ and the produced gluon are denoted
by $P$ and $k$. 
${\cal F}(x,q_{\perp })$ is the unintegrated gluon 
distribution. The production amplitude $\psi _{J/\Psi\,g 
}^{c_{2}c_{1}}(x_{1},x_{2},q_{1\perp },q_{2\perp },P,k)$ is 
factorized in a hard part which describes the production of the $ 
q\bar q$ pair and the gluon  and an amplitude describing the binding of this pair 
into a physical charmonium state. The explicit form of the hard part of
the
amplitude is given by the formulas from \cite{chi} supplemented with 
an usual additional gluon production vertex (figs. \ref{graphs} a, \ref{graphs}b). 
The formalisms used to describe the binding of the $q \bar q$ pair into
a bound state as well as the unintegrated gluon distribution are
the same as in \cite{chi}.



\begin{figure}[h] 
\centerline{\epsfig{file=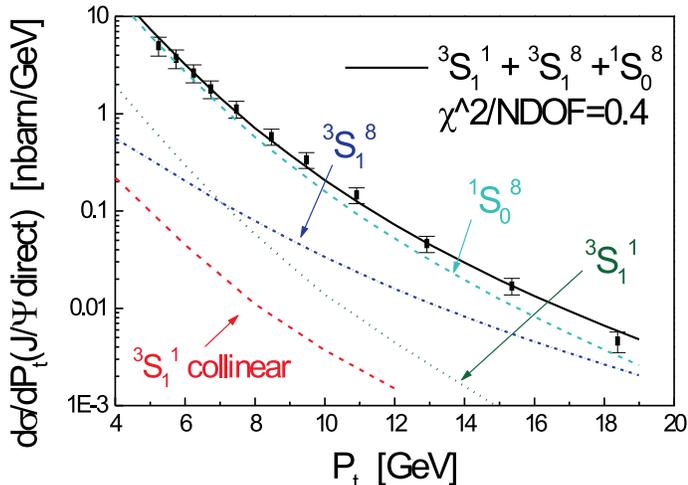,width=9cm}} 
\caption{A comparison of the CS contribution in the collinear
factorization
and the CS + CO contributions in the $k_\perp$-factorization to direct J/$\Psi$ production} 
\label{jpsi} 
\end{figure}

Our results for the direct $J/\Psi $ production cross section are shown in 
fig.\ref{jpsi} in comparison to the Tevatron data \cite{data} and the color singlet NLO QCD 
calculations in collinear factorization \cite{CLI,CLII}. Although the color singlet 
contribution for $k_{\perp }$ factorization is substantially larger than 
the corresponding part in collinear factorization it still lies about a 
factor of 10 below the experimental results. In contrast to $\chi _{c}$ 
production this leaves more room for the color octet mechanism. Since in 
$k_{\perp }$ factorization there is no general need for an additional 
outgoing perturbative gluon in order to have some non-vanishing $P_{\perp 
J/\Psi }$ we include the color octet contribution using the NLLA  
effective $q\overline{q}$ production vertex.

A fit of the uncalculable color octet 
parameters $\langle 0|O_{8}^{J/\Psi }(^{1}S_{0})|0 \rangle$, $\langle 0|O_{8}^{J/\Psi 
}(^{3}S_{1})|0\rangle$ and $\langle 0|O_{8}^{J/\Psi }(^{3}P_{0})|0\rangle$ which contribute in 
the lowest order of the NRQCD velocity expansion gives a good decription of 
the data (see figure \ref{jpsi}). As in the $\chi _{c}$ production case the $ 
^{3}S_{1}^{8}$ part is smaller than for collinear factorization. 
This is again due to the flatter slope of 
the contribution which comes from the additional parts of the 3-gluon-vertex 
in the effective NLLA $q\overline{q}$ production vertex. Since the slopes of 
the $^{1}S_{0}^{8}$ and $^{3}P_{J=0,1,2}^{8}$ parts are very similar the fit 
cannot distinguish between these two, therefore only a certain linear 
combination  
\[ 
M_{8}=\frac{\langle 0|O_{8}^{J/\Psi }(^{3}P_{0})|0\rangle}{m_{c}^{2}}+\frac{
\langle 0|O_{8}^{J/\Psi }(^{1}S_{0})|0\rangle}{R} 
\] 
of these parameters can be extracted \cite{CLI,CLII}. In our case the factor $R$ lies 
between $6$ (low $P_{\perp }$)  and $4.5$ (high $P_{\perp }$). Taking only 
the $^{1}S_{0}^{8}$ or the $^{3}P_{J=0,1,2}^{8}$ contributions we obtain two 
slightly different results (for the color singlet contribution we take 
parameters from \cite{eichten}):

\vspace*{.5cm}
\begin{tabular}{|l|l|l|}
\hline 
 & 
\begin{tabular}{l}
only $^{1}S_{0}^{8}$ contr.\\ $\langle O_8^{J/\Psi}(^{3}P_{J})\rangle=0$
\end{tabular}
& 
\begin{tabular}{l}
only
$^{3}P_{J}^{8}$ contr.\\$\langle O_8^{J/\Psi}(^{1}S_0)\rangle=0$
\end{tabular}
\\
\hline  
$\langle 0|O_{8}^{J/\Psi }(^{3}S_{1})|0\rangle$ & $3.2(\pm 1.2)\cdot 10^{-4}$ & $5(\pm 
1.2)\cdot 10^{-4}$ \\
\hline 
$M_{8}$ ($R=5$) & $1.4(\pm 0.1)\cdot 10^{-2}$ & $1.2(\pm 0.1)\cdot 10^{-2}$ \\
\hline
total $\chi ^{2}/$NDOF & $0.42$ & $0.53$\\
\hline 
\end{tabular}

\vspace*{.5cm} 
The errors are only statistical.  

$\mbox{}$


On the basis of these results we conclude that it is impossible to  
describe the data for direct $J/\Psi$ production
entirely by the CS contribution.
In our case the discrepancy between the CS contribution and the data 
is substantially smaller than for the NLO  
collinear factorization.  
 
The results allow for a good description in the framework of the COM.
We performed new fits of CO matrix elements.  
As a result, the value of $\langle 0|O_8^{J/\Psi}(^3 S_1)|0\rangle$ is reduced by 
a factor of $\approx 30$ in comparison with the analysis in the framework 
of collinear factorization. However, if this  
matrix element is put exactly to zero, the quality of the fit is much
worse.  



Let us note that from the strong reduction of the 
$\langle 0|O_8^{J/\Psi}(^3 S_1)|0\rangle$ matrix 
element in comparison to the collinear approximation 
it doesn't follow that the same happens in the case of
$\langle 0|O_8^{J/\Psi}(^1 S_0)|0\rangle$ and  
$\langle 0|O_8^{J/\Psi}(^3 P_0)|0\rangle$. Indeed we find that
the linear combination $M_8$, which we can extract from a fit to the
data, is of same order of magnitude as the analogous combination
in the collinear case. Taking this together we see that our
larger cross sections primarily give a reduction
of the $^3 S_1 ^8$ contribution and do not
lead to a uniform decrease of all color octet matrix elements.
Now since $^1 S_0 ^8$ and $^3 P_J ^8$
contributions to $J/\Psi$ enter in a different combination
 into the photoproduction cross
section (while the contribution of
$^3 S_1 ^8$ is negligible), 
we cannot make definite predictions about the impact of 
our $k_t$-factorization calculation on $J/\Psi$ photoproduction.
Besides this there is of course the need for considering higher
order subprocesses in order to determine the color octet matrix elements 
more accurately.

In the context of photoproduction
we would like to comment on the article \cite{cacciari}
in which the photoproduction of charmonia in collinear factorization 
is studied. 
There the color octet contributions 
are taken into account,
using for the CO matrix elements 
the values obtained in the collinear approach by 
Cho and Leibovich \cite{CLII} for the
hadroproduction of quarkonia. 
The results of \cite{cacciari} for
$p_{\perp J/\Psi} \geq 1 GeV$ (see fig. 4 in \cite{cacciari}) 
suggest that our 
values for the octet matrix elements support
the good description of the data shown there. The only
discrepancy occurs in the region of small $p_{\perp J/\Psi}$ 
(see \cite{cacciari}, fig.3). But as it was shown in the recent
work by Braaten et al \cite{BFL} the naive use of the color 
octet mechanism doesn't allow to describe the data in the small
$p_\perp$ region. The reason for this could be a stronger 
interplay of the soft-gluon emission from the hard 
$q \bar{q}$ subprocess with the soft gluon radition described by the color
octet mechanism.


The smallness of the color octet matrix element
$\langle 0|O_8^{J/\Psi}(^3 S_1)|0\rangle$
is, however, quite promising from the point of view 
of $J/\Psi$ polarization. 
Please recall, that in the collinear factorization approach 
this matrix element provides the dominant contribution to the cross-section 
through the gluon fragmentation subprocess. 
As soon as the gluon
is almost on-shell, it has a strong transverse polarization 
which should result in a strong transverse polarization of 
$J/\Psi$, in disagreement with the experimental data \cite{pol,braaten}. 
In contrast, in our approach this fragmentation mechanism is
suppressed as it leads to a wrong $P_\perp$-dependence. 
Consequently the qualitative origin 
 of the transverse polarization in the collinear approach 
is absent in the $k_\perp$-factorization approach.
In contrast the longitudinal polarization of the colliding gluons in
the
$k_\perp$-factorization approach leads to a vanishing 
projection of the angular momentum of the produced $q\bar q$-pair on
the collision axis resulting in
particular in a longitudinal polarization of produced $\chi_c$'s.

Finally let us comment on the recent articles by F. Yuan and
K.-T. Chao \cite{yuan,YuanPol,YuanChi} who studied polarized and
unpolarized charmonium production based on 
our approach \cite{chi,open}.
In \cite{yuan} they performed independently analogous 
calculations as in the present paper
and found very similar numerical results.
They have shown \cite{YuanPol} 
that the $J/\Psi$'s are in fact nearly
unpolarized, 
which confirms 
our expectation on the absence of strong transverse
polarization in charmonium hadroproduction.
Furthermore they demonstrated that $\chi_c$ states
are predominantly longitudinally polarized \cite{YuanChi}.



We acknowledge the discussion with A. Tkabladze.
This work was supported by DFG and BMBF.
O. T. was also supported in part by RFFI grant 00-02-16696.

\end{document}